\documentclass{elsart1p}

\usepackage{graphicx}

\usepackage{amssymb}
\usepackage{amsmath}

\usepackage{lineno}

\linenumbers
\journal{Physica D}
\begin{document}

\begin{frontmatter}


\title{Collisions of a light bullet with kinks and standing breathers 
in the two-dimensional sine-Gordon equation}

\author{M. V. Pato\corauthref{cor1}}
\ead{migpato@gmail.com}
\corauth[cor1]{Corresponding author.}
\address{Dep. F\'isica, Instituto Superior T\'ecnico, Universidade T\'ecnica de Lisboa}

\author{P. Bicudo}
\ead{bicudo@ist.utl.pt}
\address{CFTP, Dep. F\'{\i}sica, Instituto Superior T\'ecnico, Av. Rovisco Pais, 1049-001 Lisboa, Portugal}





\begin{abstract}
The (2+1)-dimension Klein-Gordon generalised equation is numerically solved through the finite differences method. Only the sine-Gordon case is focused: kink and antikink solutions are obtained in cartesian coordinates and evidence of interaction in kink-kink collision is looked for in propagation velocity. Then the change of shape in light bullet solutions is quantified during propagation and in head-on collision. Lastly, the robustness of light bullets is verified in head-on collisions with kink, antikink, standing kink and standing breather. A 30$^{\circ}$-collision between a light bullet and a standing kink is simulated as well.
\end{abstract}

\begin{keyword}
sine-Gordon equation \sep soliton \sep light bullet \sep kink \sep standing breather
\PACS 02.30.Jr \sep 02.70.Bf \sep 03.75.Lm \sep 42.81.Dp
\end{keyword}
\end{frontmatter}


\section{Introduction}
\par The Klein-Gordon generalised equation is written as
\begin{equation}\label{eqKGgen}
\nabla^{2} \phi-\frac{1}{c^2}\frac{\partial^2\phi}{\partial t^2}=F(\phi)
\end{equation}
where $F$ is an arbitrary function of $\phi$. The well-known wave equation is obtained with $F(\phi)=0$, while $F(\phi)=\left(\frac{mc}{\hbar}\right)^2\phi$ leads to Klein-Gordon equation that describes a particle of mass $m$ and spin 0 in relativistic quantum mechanics. Both cases are linear and only the latter is dispersive \cite{SGS}. An important nonlinear case is characterised by $F(\phi)=\left(\frac{mc}{\hbar}\right)^2 sin\phi$:
\begin{equation}\label{eqSG}
\nabla^{2} \phi-\frac{1}{c^2}\frac{\partial^2\phi}{\partial t^2}=\left(\frac{mc}{\hbar}\right)^2 sin\phi
\end{equation}

\par Equation \eqref{eqSG} is the so-called sine-Gordon equation and allows soliton-like solutions. This type of solutions is increasingly important in the description of (at least partly) particle-like objects \cite{DiGarbo,pion}, which is easy to understand since solitons present very localised momentum and their shape consistency results in effective transportation of energy. For instance, the $\pi$-mesons are quark-antiquark bound states and some works (e.g. \cite{pion}) have been able to estimate pion mass values in good agreement with the experiment by describing pions as breather-like solutions of sine-Gordon equation. In other words, the quark may be seen as a soliton (e.g. kink) and the antiquark as an antisoliton (e.g antikink) so that $\pi$ is a bound state of the two. Another example is the Josephson junction, where the phase difference between the electronic densities of the superconductors may be modeled by sine-Gordon equation which previews the vortex dynamics in type II superconductors \cite{Sobolev,Wallraff}. The sine-Gordon equation finds also application in classical systems such as coupled pendula \cite{artgall} and, finally, light bullet solutions may be identified with optical pulses propagating in different media \cite{Povich,Xin}.

\par In this paper, section \ref{2} briefly describes the numerical method used, while section \ref{3} presents the results of the simulations and is organised as follows. In \ref{31} the propagation and collision of kink-like solitons is analysed and in \ref{32} an analogous study is performed for light bullet solutions. Section \ref{33} characterises the simulation of the collision between light bullets and other sine-Gordon solutions and, finally, in section \ref{4} the most important remarks are drawn.

\section{Numerical method}\label{2}
\par In bidimensional cartesian coordinates, the (2+1)-dimension Klein-Gordon generalised equation becomes $\frac{\partial^2\phi}{\partial x^2}+\frac{\partial^2\phi}{\partial y^2}-\frac{1}{c^2}\frac{\partial^2\phi}{\partial t^2}=F(\phi)$. With the normalisation substitutions $x=\frac{\hbar}{mc}x'$, $y=\frac{\hbar}{mc}y'$ and $t=\frac{\hbar}{mc^2}t'$ and loosing the primes, the following equation results:
\begin{equation}\label{eqKGgenCart}
\frac{\partial^2\phi}{\partial x^2}+\frac{\partial^2\phi}{\partial y^2}-\frac{\partial^2\phi}{\partial t^2}=\left(\frac{\hbar}{mc}\right)^2 F(\phi)
\end{equation}

\par In order to numerically solve \eqref{eqKGgenCart}, the finite differences method is implemented in \textit{Mathematica 5.0} \textregistered. A rectangular domain $\left\{\left(t,x,y\right): t\epsilon\left[t_1,t_{n_t}\right] \wedge x\epsilon\left[x_1,x_{n_x}\right] \wedge y\epsilon\left[y_1,y_{n_y}\right] \right\}$ is chosen and an $n_t\cdot n_x\cdot n_y$-point space-time grid is created with uniform steps $\Delta t=\frac{t_{n_t}-t_1}{n_t-1}$, $\Delta x=\frac{x_{n_x}-x_1}{n_x-1}$ and $\Delta y=\frac{y_{n_y}-y_1}{n_y-1}$. Second-order derivatives are approximately computed through centered second-order differences.

\par Initial conditions are given by the definition of $\phi$ and $\frac{\partial\phi}{\partial t}$ at time $t_1$. As for boundary conditions, for all simulations presented here fixed conditions are used along $y$-direction and periodic ones along $x$-direction.

\par One must take into account the convergence conditions of the method, $\frac{c^2 \Delta t^2}{\Delta x^2}\leq\frac{1}{2}$ and $\frac{c^2 \Delta t^2}{\Delta y^2}\leq\frac{1}{2}$, and, in the case of space-periodic solutions, the spatial grid's resolution has to be adequate: a solution with spatial periodicity through (a,b)-direction and with period $\delta$ should be represented with some - say, 3 - points per period, that is, $\Gamma=\frac{\delta}{a\Delta x+b\Delta y}\geq 3$. Otherwise, meaningless results may be obtained.

\par The whole numerical approach reffers to Klein-Gordon generalised equation. However, in this paper, only the sine-Gordon case is studied; thus, from now on $F(\phi)=\left(\frac{mc}{\hbar}\right)^2 sin\phi$. Future developments may consider any other case using the same numerical method with the corresponding definition of $F$.

\section{Results}\label{3}
\subsection{Kink-like solitons}\label{31}
\par Theoretical approaches to (1+1)-dimension sine-Gordon equation \cite{Infeld,Whitham} allow one to write an 1-soliton analytical solution in the (2+1)-dimension case as
\begin{equation}\label{1sol}
\phi\left(t,\stackrel{\rightarrow}{r}\right)=4 arctg\left(Sign(m)\cdot e^{\frac{\stackrel{\rightarrow}{d}\cdot\left(\stackrel{\rightarrow}{r}-\stackrel{\rightarrow}{r_0}\right)-\beta t}{\sqrt{1-\beta^2}}}\right)
\end{equation}
where $\beta=\frac{\sqrt{m^2-1}}{m}$, $\left|m\right|\geq 1$, $\stackrel{\rightarrow}{d}$ is the normalised propagation direction and $\stackrel{\rightarrow}{r_0}$ the position of the center point at $t=0$. 

\par
\begin{figure}[h]
\centering 
\includegraphics[scale=0.43]{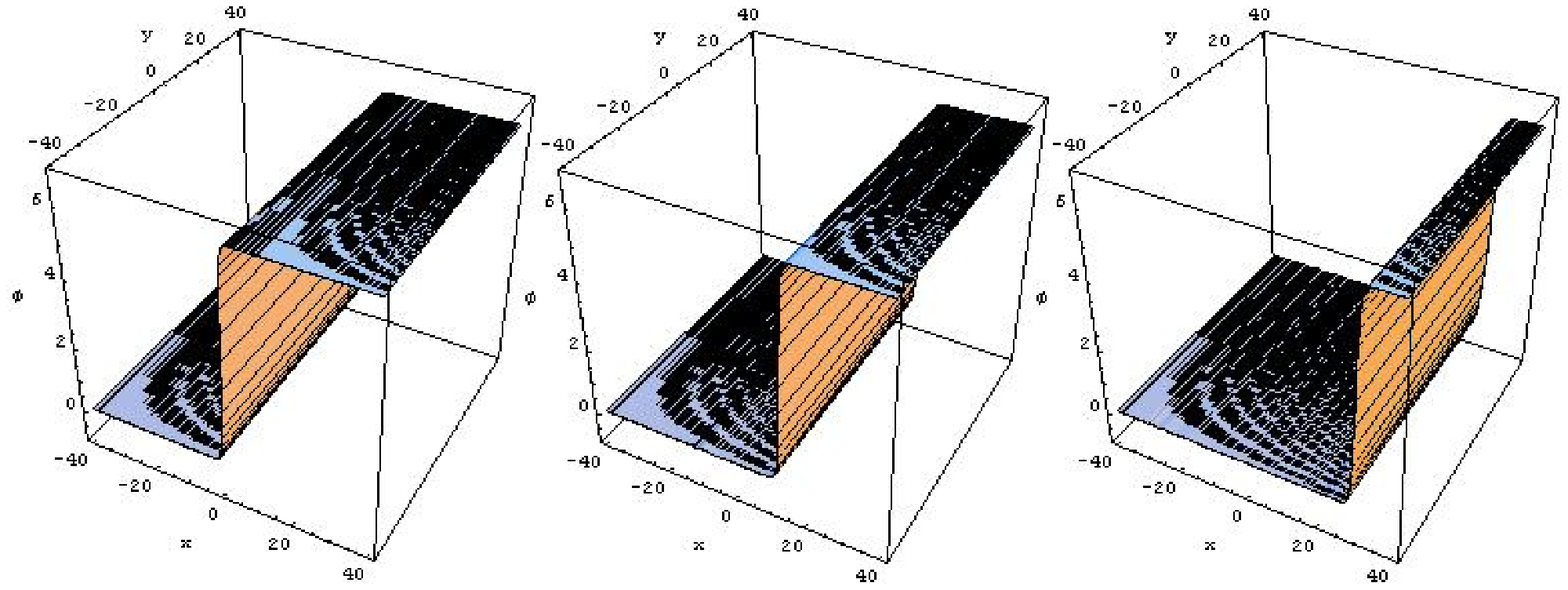}
\caption{Propagation of a kink with $m=2$, $\stackrel{\rightarrow}{d}=\stackrel{\rightarrow}{e_x}$ and $\stackrel{\rightarrow}{r_0}=\stackrel{\rightarrow}{0}$ within an $800\cdot401\cdot2$-point grid defined in $\left\{\left(t,x,y\right): t\epsilon\left[0,100\right] \wedge x\epsilon\left[-40,40\right] \wedge y\epsilon\left[-40,40\right] \right\}$. In the sequence of shots time flows from left to right.}
\label{fig2}
\end{figure}

\begin{figure}[h]
\centering 
\includegraphics[scale=0.43]{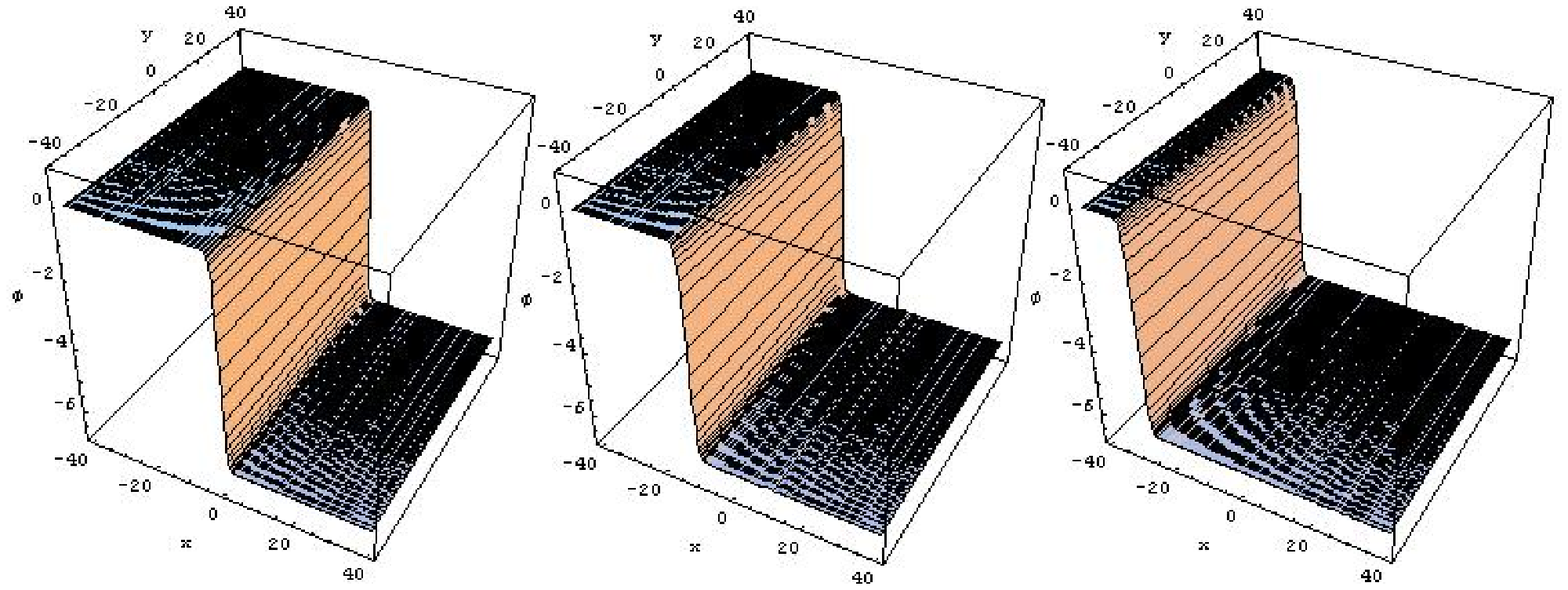}
\caption{Propagation of an antikink with $m=-2$, $\stackrel{\rightarrow}{d}=\stackrel{\rightarrow}{e_x}$ and $\stackrel{\rightarrow}{r_0}=\stackrel{\rightarrow}{0}$ within an $800\cdot401\cdot2$-point grid defined in $\left\{\left(t,x,y\right): t\epsilon\left[0,100\right] \wedge x\epsilon\left[-40,40\right] \wedge y\epsilon\left[-40,40\right] \right\}$.}
\label{fig3}
\end{figure}

 The case $m>1$ corresponds to the so-called \textit{kink}, an ascending step propagating along $\stackrel{\rightarrow}{d}$ with velocity $\beta$. An example is presented on figure \ref{fig2}. Analogously, $m<-1$ corresponds to an \textit{antikink}, a descending step propagating along $\stackrel{\rightarrow}{d}$ with velocity $-\beta$ (see figure \ref{fig3}). Both kink and antikink behave like solitons since they maintain their shape while propagating. Particular solutions are the cases $m=1$ (\textit{standing kink}) and $m=-1$ (\textit{standing antikink}) which are stationary.

\subsubsection{Propagation}\label{311}

\par
\begin{figure}[h]
\centering 
\includegraphics[scale=0.43]{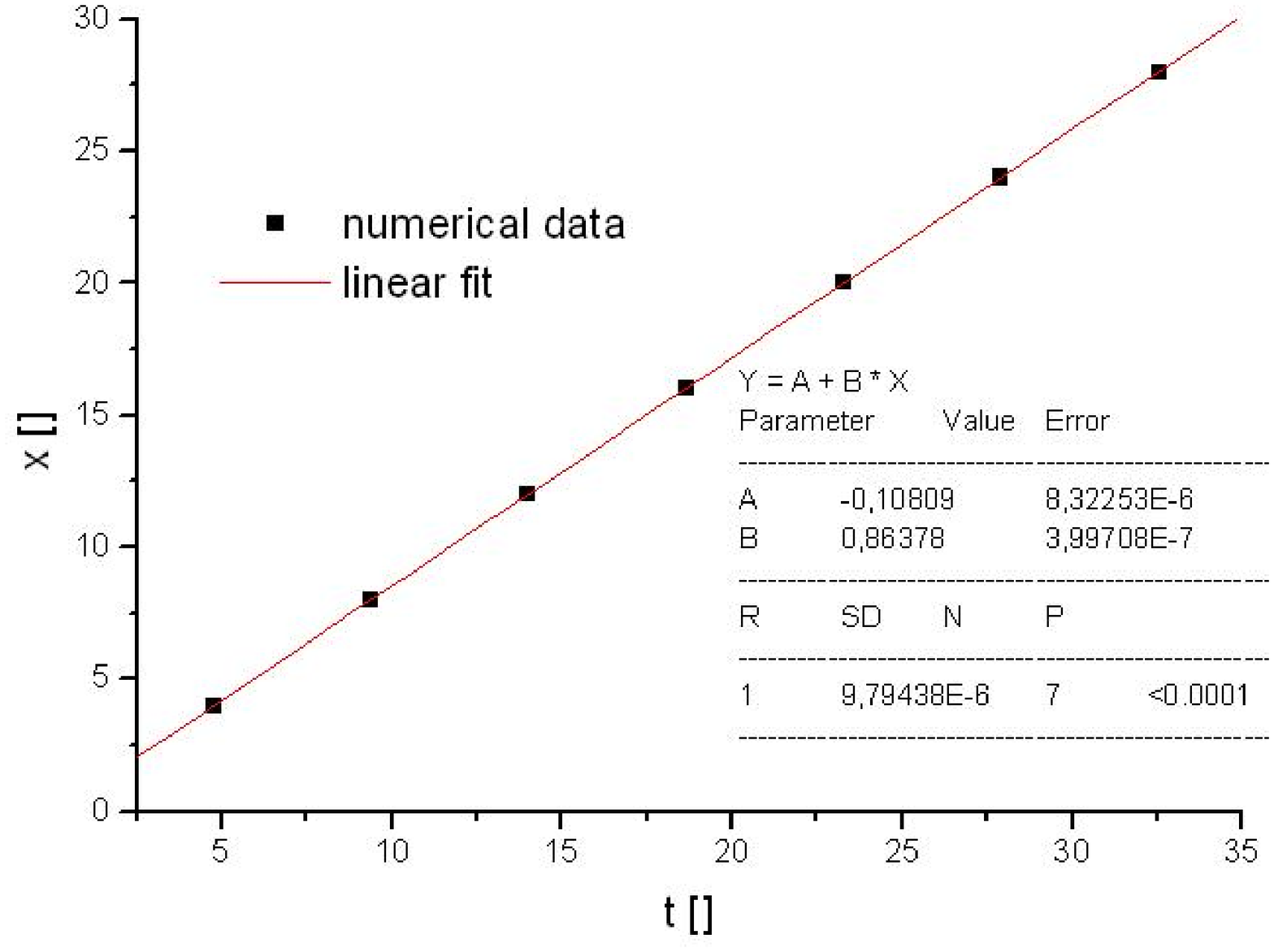}
\caption{Linear fit of pairs $\left\{time,position\right\}$ of the center point of a kink with $m=2$, $\stackrel{\rightarrow}{d}=\stackrel{\rightarrow}{e_x}$ and $\stackrel{\rightarrow}{r_0}=\stackrel{\rightarrow}{0}$ propagating within an $800\cdot401\cdot2$-point grid defined in $\left\{\left(t,x,y\right): t\epsilon\left[0,100\right] \wedge x\epsilon\left[-40,40\right] \wedge y\epsilon\left[-40,40\right] \right\}$.}
\label{fig4}
\end{figure}

 The dependence on the $m$ parameter of the kink solution is studied. Regardless of the value of $m$, a nonlinear trail dued to numerical reasons is always present in the propagation of the kink. As $m$ rises, the trail becomes more significant, which may be explained by the steepening of the step. Indeed, $\phi\propto arctg\left(e^{\frac{x}{\sqrt{1-\beta^2}}}\right)=arctg\left(e^{|m|x}\right)$ $\left(\stackrel{\rightarrow}{d}=\stackrel{\rightarrow}{e_x}, \stackrel{\rightarrow}{r_0}=\stackrel{\rightarrow}{0}\right)$. For $m=7$ or higher, the kink is destabilised and no proper propagation is achieved. Another important $m$-dependent feature is the propagation velocity. The theoretical motion is given by $x=x_0+\beta t$ $\left(if \stackrel{\rightarrow}{d}=\stackrel{\rightarrow}{e_x}\right)$; so, the velocity should be constant and equal to $\beta$. To determine this velocity numerically, several pairs of time and position of the center point are registered and a linear fit is applied using \textit{Origin 5.0} \textregistered  - an example is presented in figure \ref{fig4}. The results for $m=\left\{1.5,2.0,2.5,3.0\right\}$ $\left(using \stackrel{\rightarrow}{d}=\stackrel{\rightarrow}{e_x}, \stackrel{\rightarrow}{r_0}=\stackrel{\rightarrow}{0}\right)$ are presented in table \ref{table1}. It is obvious that greater values of $m$ lead to less precision and exactness on the determination of the velocity. Taking into account the last column of table \ref{table1}, the most favourable situation occurs when $m=2.0$. Therefore, the next sections will use this kind of kinks (and the corresponding antikinks with $m=-2.0$) as they are less disturbed by numerical errors. Moreover, the fitted values of $x_0$ are close to 0, which means the law $x=x_0+\beta c t$ is being followed $\left(\stackrel{\rightarrow}{r_0}=\stackrel{\rightarrow}{0}\right)$.
\par It is interesting to note that all velocities in table \ref{table1} are below 1 ($c=1$), which is the typical velocity of a wave in the case $F(\phi)=0$. As $m\rightarrow \infty$, the velocity of the kink tends to 1 since $\beta\rightarrow 1$.

\begin{table}[h]
\begin{center}
\begin{tabular}{c|c|c|c|c}
\hline\hline  m [] & $x_0^{(num)}$ [] & $\beta ^{(num)}$ [] & $\beta ^{(teo)}$ [] & $\delta_{\beta}$ [\%] \\ \hline 
1.5 & -0.093 $\pm$ 0.000 & 0.743 $\pm$ 0.000 &  0.745 & 0.28 \\ \hline
2.0 & -0.108 $\pm$ 0.000 & 0.864 $\pm$ 0.000 & 0.866 & 0.26 \\ \hline
2.5 & -0.081 $\pm$ 0.028 & 0.908 $\pm$ 0.001 &  0.917 & 0.97 \\ \hline
3.0 & 0.017 $\pm$ 0.039 & 0.922 $\pm$ 0.002 &  0.943 & 2.25 \\
\hline \hline
\end{tabular}\caption{Determination of the propagation velocity of kinks with $m=\left\{1.5,2.0,2.5,3.0\right\}$, $\stackrel{\rightarrow}{d}=\stackrel{\rightarrow}{e_x}$ and $\stackrel{\rightarrow}{r_0}=\stackrel{\rightarrow}{0}$ propagating within an $800\cdot401\cdot2$-point grid defined in $\left\{\left(t,x,y\right): t\epsilon\left[0,100\right] \wedge x\epsilon\left[-40,40\right] \wedge y\epsilon\left[-40,40\right] \right\}$. The last column shows the relative difference between $\beta^{(num)}$ and $\beta^{(teo)}$: $\delta_{\beta}=\frac{\left|\beta^{(num)}-\beta^{(teo)}\right|}{\beta^{(teo)}}\cdot 100$.}\label{table1}
\end{center}
\end{table}

\subsubsection{Collision}

\par
\begin{figure}[h]
\centering 
\includegraphics[scale=0.43]{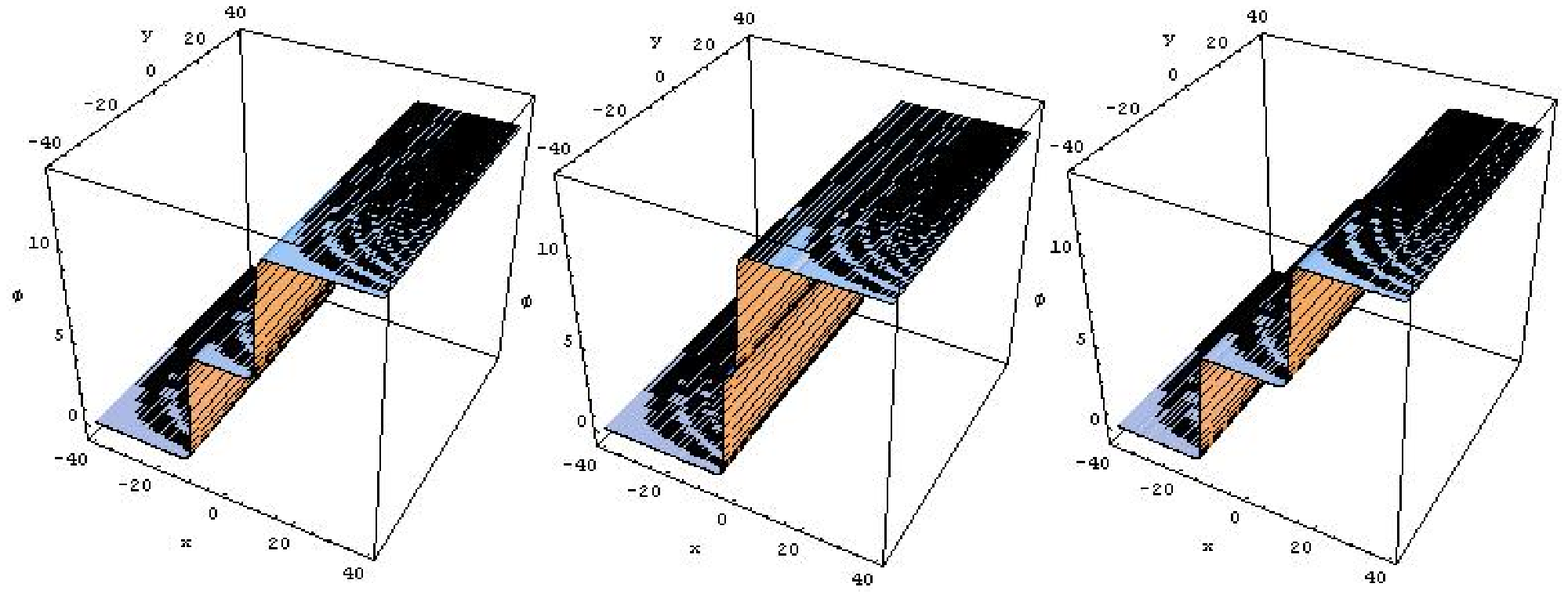}
\caption{Kink-kink collision. Both kinks have $m=2$ and  $\stackrel{\rightarrow}{d}=\stackrel{\rightarrow}{e_x}$; the kink propagating left-to-right (K$\rightarrow$) has $\stackrel{\rightarrow}{r_0}=(-10,0)$, while the one propagating right-to-left (K$\leftarrow$) has $\stackrel{\rightarrow}{r_0}=(10,0)$. An $800\cdot401\cdot2$-point grid is defined in $\left\{\left(t,x,y\right): t\epsilon\left[0,100\right] \wedge x\epsilon\left[-40,40\right] \wedge y\epsilon\left[-40,40\right] \right\}$.}
\label{fig5}
\end{figure}

 The collision between kink-like solitons is achieved by superposing them. However, in general, this procedure is 
not legitimate since the sine-Gordon equation \eqref{eqSG} is nonlinear. The condition for the coexistence
of a pair of legitimate kink-like solitons is 
\begin{equation}\label{superpose}
sin(\phi_1+\phi_2)=sin\phi_1+sin\phi_2
\end{equation}
where $\phi_1$ and $\phi_2$ are solutions of \eqref{eqSG}. Thus, two kink-like solitons may be superposed if their 
steps are sufficiently separated, because in that case on each grid point one of the two solutions is a multiple 
of $2\pi$, which means \eqref{superpose} is verified. In this way, the kink-kink collision, shown in figure \ref{fig5}, 
can be easily started and studied. Only the collision of kinks with symmetric velocity and same $|m|$ is simulated. 
Future works may take into account other situations.

\par Figure \ref{fig5} shows that the kinks emerge intact (shape-wise) from collisions, which is typical of a soliton's behaviour. However, as they are nonlinear objects, they interact with each other during collisions. Evidence of this interaction is looked for in solitons velocity. The determination of the velocity follows the procedure explained in section \ref{311}, but picking pairs of time and position only after the interaction. The results obtained for each of the kinks are presented in table \ref{table2}. Comparing these results with those referring to propagation only (table \ref{table1}, line with $m=2.0$), one understands that the value of $\beta$ does not change significantly, while $x_0$ is now clearly above 0. This means that the interaction speeds up each one of the kinks, but does not alter their velocity afterwards.

\begin{table}[h]
\begin{center}
\begin{tabular}{c|c|c}
\hline\hline   & $x_0^{(num)}$ [] & $\beta^{(num)}$ [] \\ \hline
K$\rightarrow$ & 0.703 $\pm$ 0.000 & 0.864 $\pm$ 0.000 \\ \hline
K$\leftarrow$ & 0.703 $\pm$ 0.000 & 0.864 $\pm$ 0.000 \\ \hline \hline
\end{tabular}
\caption{Determination of the propagation velocity of the kinks in the kink-kink collision referenced by figure \ref{fig5}.}\label{table2}
\end{center}
\end{table}

\subsection{Light bullets}
\label{32}
\par The \textit{light bullet} \cite{Povich,Xin} represents a well-localised two-dimensional moving pulse. Its previewed space-time dependence is
\begin{eqnarray}\label{lb}
\phi\left(t,\stackrel{\rightarrow}{r}\right)=&sin&  \left(\gamma \stackrel{\rightarrow}{d} \cdot \left(\stackrel{\rightarrow}{r}-\stackrel{\rightarrow}{r_0}\right)+\omega t\right) \cdot \nonumber \\
& \cdot e&^{-\frac{1}{4\sigma^2}\left(\left[\stackrel{\rightarrow}{d} \cdot \left(\stackrel{\rightarrow}{r}-\stackrel{\rightarrow}{r_0}\right)\right]^2+\left[\stackrel{\rightarrow}{u} \cdot \left(\stackrel{\rightarrow}{r}-\stackrel{\rightarrow}{r_0}\right)\right]^2\right)}
\end{eqnarray}
where $\omega=\sqrt{1+\gamma^2}$ and $\stackrel{\rightarrow}{u}$ is a normalised direction perpendicular to $\stackrel{\rightarrow}{d}$.

\subsubsection{Propagation}\label{321}
\par
\begin{figure}[h]
\centering 
\includegraphics[scale=0.43]{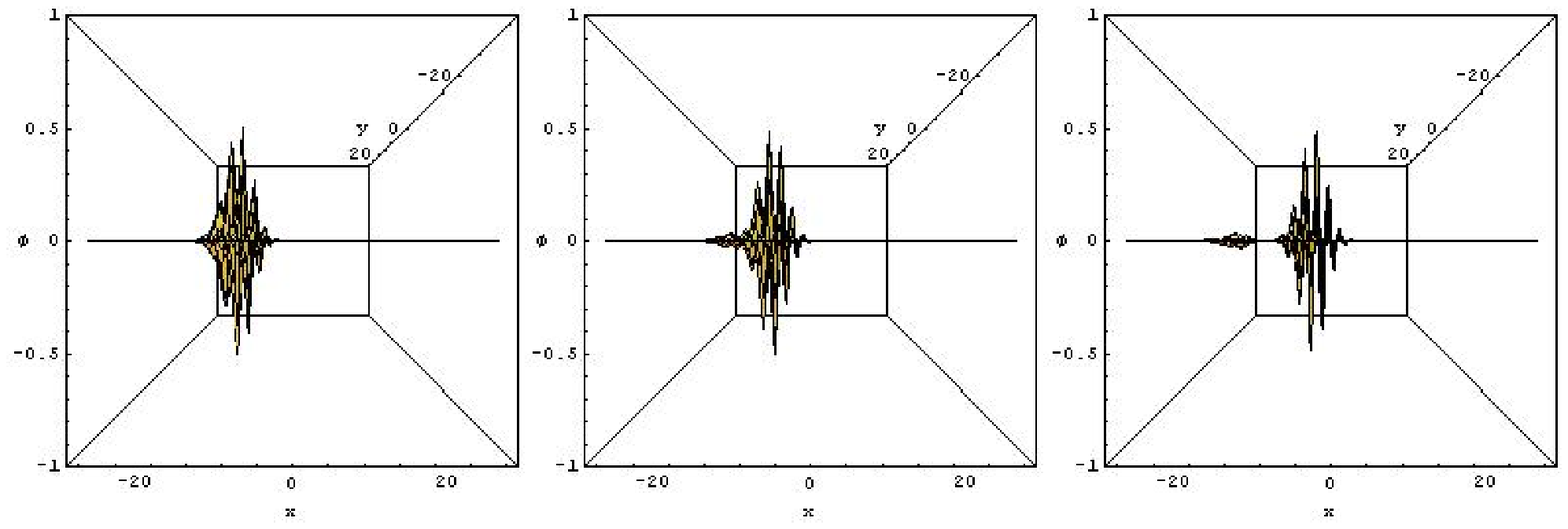}
\caption{Propagation of a light bullet with $\left\{\gamma,\sigma\right\}=\left\{2.0,2.5\right\}$, $\stackrel{\rightarrow}{r_0}=(-15,0)$, $\stackrel{\rightarrow}{d}=\stackrel{\rightarrow}{e_x}$ and $\stackrel{\rightarrow}{u}=\stackrel{\rightarrow}{e_y}$. An $800\cdot151\cdot15$-point grid is defined in $\left\{\left(t,x,y\right): t\epsilon\left[0,100\right] \wedge x\epsilon\left[-30,30\right] \wedge y\epsilon\left[-30,30\right] \right\}$. The images are the projection of the solution in the xz plane. In this case, $\Gamma=7.85$.}
\label{fig11}
\end{figure}

 The evolution of a single light bullet is represented in figure \ref{fig11}. There is no significant modification of shape - light bullets behave indeed as solitons. Nevertheless, a nonlinear trail is formed and the $\sigma$-value of the envelope function changes during propagation. To quantify the former, one compares the integral of $\phi$ in the trail region to that in the pulse region:
\begin{displaymath}
\epsilon=\frac{\int_{trail} \phi dx dy}{\int_{pulse} \phi dx dy}
\end{displaymath}
The higher $\epsilon$, the more important the trail is in comparison with the soliton. It is important to note that the trail is dued not only to numerical reasons but also to the fact that \eqref{lb} is not an analytical solution of \eqref{eqSG}, unlike the case of the kink-like solitons. The latter effect is evaluated fitting the numerical values of $\phi$ after propagation to
\begin{displaymath}
a_0+a_1e^{-\frac{1}{4 {a_2}^2}\left(\left[\stackrel{\rightarrow}{d} \cdot \left(\stackrel{\rightarrow}{r}-\stackrel{\rightarrow}{a_3}\right)\right]^2+\left[\stackrel{\rightarrow}{u} \cdot \left(\stackrel{\rightarrow}{r}-\stackrel{\rightarrow}{a_3}\right)\right]^2\right)}
\end{displaymath}
where sometimes not all parameters are free to vary for convergence purposes.

\par One can now study $\epsilon$ and $a_2$ as functions of $\gamma$ and $\sigma$ and then choose the most 
favourable pair $\left\{\gamma,\sigma\right\}$ to perform collisions between light bullets. 
Tables \ref{table4} and \ref{table5} present the results of these simulations. Although the 
case $\left\{\gamma,\sigma\right\}=\left\{1.0,5.0\right\}$ is the one with less significant 
trail, it leads to a strong deviation of $\sigma$ after propagation. The pair of parameters 
$\left\{\gamma,\sigma\right\}=\left\{2.0,2.5\right\}$ seems to be the one in which less 
numerical errors occur; therefore, these are the values used in the next sections.

\begin{table}[!h]
\begin{center}
\begin{tabular}{c|c|c|c}
\hline\hline  $\sigma$ [] $\backslash$ $\gamma$ []  & 1.0     & 2.0         & 3.0   \\
\hline        1.0                                   & 0.0402  & 0.0439      & 0.0447\\
\hline        2.5                                   & 0.0448  & 0.0430      & 0.0442\\
\hline        5.0                                   & 0.0380  & 0.0487      & 0.0461\\
\hline        $\Gamma$ []                           & 15.71   & 7.85        & 5.23  \\
\hline\hline
\end{tabular}\caption{Values of $\epsilon$ for light bullets with $\gamma=\left\{1.0,2.0,3.0\right\}$ and $\sigma=\left\{1.0,2.5,5.0\right\}$ propagating within an $800\cdot151\cdot15$-point grid defined in $\left\{\left(t,x,y\right): t\epsilon\left[0,100\right] \wedge x\epsilon\left[-30,30\right] \wedge y\epsilon\left[-30,30\right] \right\}$. $\Gamma$-values are also shown for each $\gamma$ used.}\label{table4}
\end{center}
\end{table}

\begin{table}[!h]
\begin{center}
\begin{tabular}{c|c|c|c}
\hline\hline  $\sigma$ [] $\backslash$ $\gamma$ []  & 1.0     & 2.0         & 3.0   \\
\hline        1.0                                   & 2.688   & 1.156       & 2.067 \\
\hline        2.5                                   & 2.704   & 2.468       & 2.573 \\
\hline        5.0                                   & 5.907   & 4.596       & 4.639 \\
\hline        $\Gamma$ []                           & 15.71   & 7.85        & 5.23  \\
\hline\hline
\end{tabular}\caption{Values of the fitting parameter $a_2$ for light bullets with $\gamma=\left\{1.0,2.0,3.0\right\}$ and $\sigma=\left\{1.0,2.5,5.0\right\}$ propagating within an $800\cdot151\cdot15$-point grid defined in $\left\{\left(t,x,y\right): t\epsilon\left[0,100\right] \wedge x\epsilon\left[-30,30\right] \wedge y\epsilon\left[-30,30\right] \right\}$. $\Gamma$-values are also shown for each $\gamma$ used.}\label{table5}
\end{center}
\end{table}

\subsubsection{Collision}

\par
\begin{figure}[h]
\centering 
\includegraphics[scale=0.4]{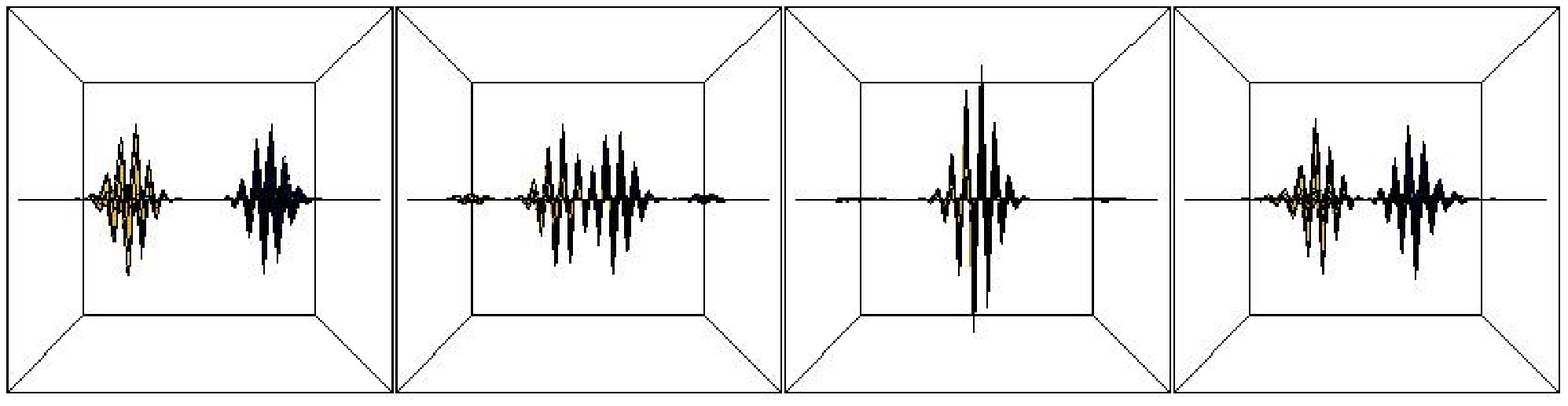}
\caption{Head-on collision between two light bullets with $\left\{\gamma,\sigma\right\}=\left\{2.0,2.5\right\}$, $\stackrel{\rightarrow}{r_{01}}=(-15,0)$ and $\stackrel{\rightarrow}{r_{02}}=(15,0)$. An $800\cdot151\cdot15$-point grid is defined in $\left\{\left(t,x,y\right): t\epsilon\left[0,100\right] \wedge x\epsilon\left[-30,30\right] \wedge y\epsilon\left[-30,30\right] \right\}$. The images are the projection of the solution in the xz plane.}
\label{fig12}
\end{figure}

 Only the collision between light bullets of $\left\{\gamma,\sigma\right\}=\left\{2.0,2.5\right\}$ is simulated. Other cases may be delt with in future approaches. The numerical method is started by superposing light bullets according to \eqref{superpose}. Figure \ref{fig12} shows a head-on collision. The solitons interact with each other and emerge essentially intact and with no change in propagation direction. However, the $\sigma$-value of the envelope function of each bullet does not behave as in the propagation case. In fact, the value of the fitting parameter $a_2$ after collision is 2.09. It seems that head-on collisions lead to a shrink of the soliton, but this conclusion must be tested by further and more complete simulations.

\subsection{Collision between light bullets and other solutions}\label{33}

\par
\begin{figure}[h]
\centering 
\includegraphics[scale=0.43]{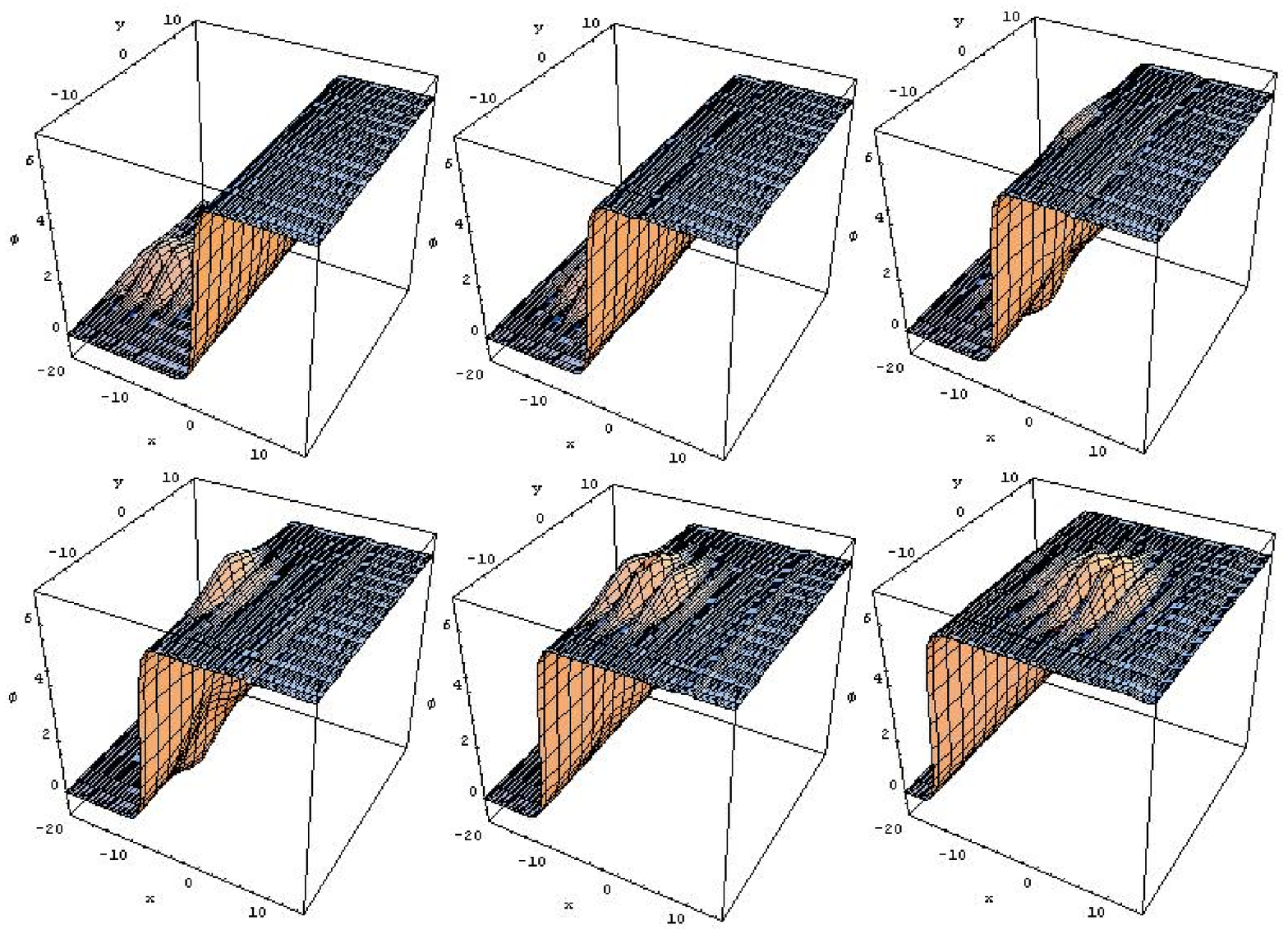}
\caption{Kink-light bullet head-on collision. The light bullet has $\left\{\gamma,\sigma\right\}=\left\{2.0,2.5\right\}$, $\stackrel{\rightarrow}{r_{01}}=(-15,0)$ and $\stackrel{\rightarrow}{d_1}=\stackrel{\rightarrow}{e_x}$, while the kink presents $m=2$, $\stackrel{\rightarrow}{r_{02}}=\stackrel{\rightarrow}{0}$ and $\stackrel{\rightarrow}{d_2}=-\stackrel{\rightarrow}{e_x}$. An $800\cdot251\cdot31$-point grid is defined in $\left\{\left(t,x,y\right): t\epsilon\left[0,100\right] \wedge x\epsilon\left[-30,30\right] \wedge y\epsilon\left[-35,35\right] \right\}$ and this domain is zoomed into $\left\{\left(t,x,y\right): t\epsilon\left[0,100\right] \wedge x\epsilon\left[-20,15\right] \wedge y\epsilon\left[-15,15\right] \right\}$. In the sequence of shots time flows from left to right and downwards. In this case, $\Gamma=13.09$.}
\label{fig13}
\end{figure}

\begin{figure}[h]
\centering 
\includegraphics[scale=0.43]{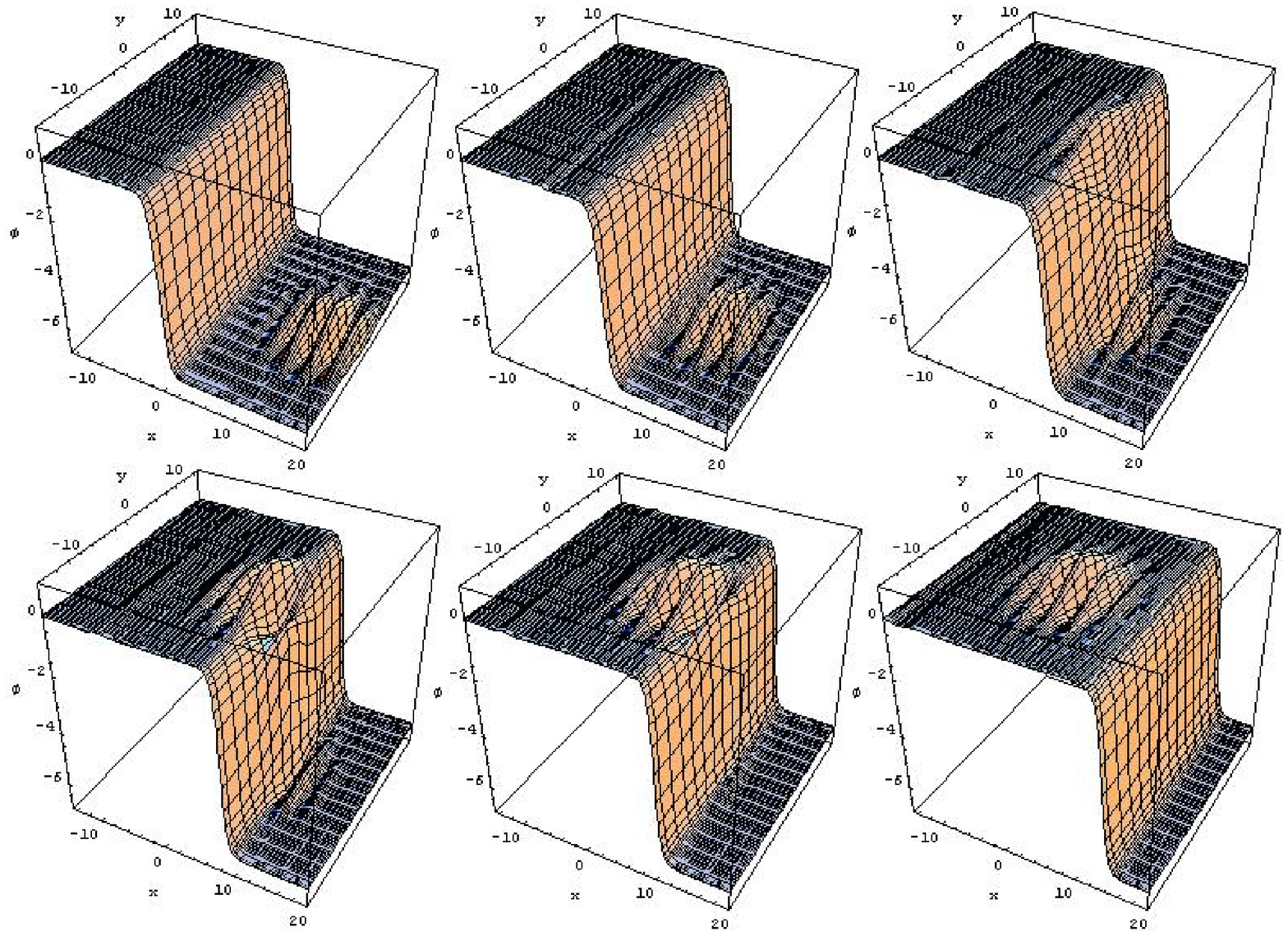}
\caption{Antikink-light bullet head-on collision. The light bullet has $\left\{\gamma,\sigma\right\}=\left\{2.0,2.5\right\}$, $\stackrel{\rightarrow}{r_{01}}=(15,0)$ and $\stackrel{\rightarrow}{d_1}=-\stackrel{\rightarrow}{e_x}$, while the antikink presents $m=-2$, $\stackrel{\rightarrow}{r_{02}}=\stackrel{\rightarrow}{0}$ and $\stackrel{\rightarrow}{d_2}=-\stackrel{\rightarrow}{e_x}$. An $800\cdot251\cdot31$-point grid is defined in $\left\{\left(t,x,y\right): t\epsilon\left[0,100\right] \wedge x\epsilon\left[-30,30\right] \wedge y\epsilon\left[-35,35\right] \right\}$ and this domain is zoomed into $\left\{\left(t,x,y\right): t\epsilon\left[0,100\right] \wedge x\epsilon\left[-15,20\right] \wedge y\epsilon\left[-15,15\right] \right\}$. In this case, $\Gamma=13.09$.}
\label{fig14}
\end{figure}

\begin{figure}[h]
\centering 
\includegraphics[scale=0.43]{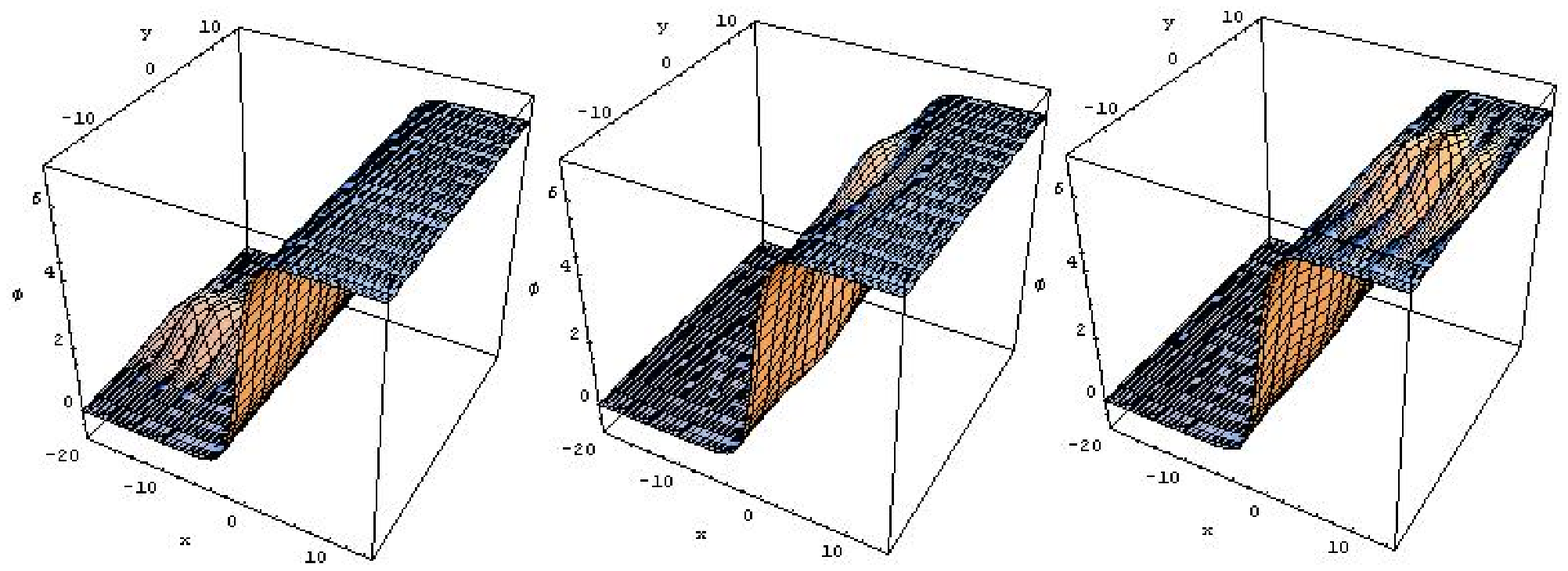}
\caption{Standing kink-light bullet head-on collision. The light bullet has $\left\{\gamma,\sigma\right\}=\left\{2.0,2.5\right\}$, $\stackrel{\rightarrow}{r_{01}}=(-15,0)$ and $\stackrel{\rightarrow}{d_1}=\stackrel{\rightarrow}{e_x}$, while the standing kink ($m=1$) presents $\stackrel{\rightarrow}{r_{02}}=\stackrel{\rightarrow}{0}$ and $\stackrel{\rightarrow}{d_2}=\stackrel{\rightarrow}{e_x}$. An $800\cdot251\cdot31$-point grid is defined in $\left\{\left(t,x,y\right): t\epsilon\left[0,100\right] \wedge x\epsilon\left[-30,30\right] \wedge y\epsilon\left[-35,35\right] \right\}$ and this domain is zoomed into $\left\{\left(t,x,y\right): t\epsilon\left[0,100\right] \wedge x\epsilon\left[-20,15\right] \wedge y\epsilon\left[-15,15\right] \right\}$. In this case, $\Gamma=13.09$.}
\label{fig15}
\end{figure}

 In order to study the interaction of light bullets with kink-like solutions, kink-, antikink- and standing kink-light bullet head-on collisions are set up and shown, respectively, in figures \ref{fig13}, \ref{fig14} and \ref{fig15}. In all three cases, the light bullet seems to emerge intact from the interaction and continues its motion with no visible change. The velocities of the light bullet in the propagation case (section \ref{321}) and in kink-light bullet head-on collision are roughly determined and no significant difference is noticed.
\par
\begin{figure}[h]
\centering 
\includegraphics[scale=0.43]{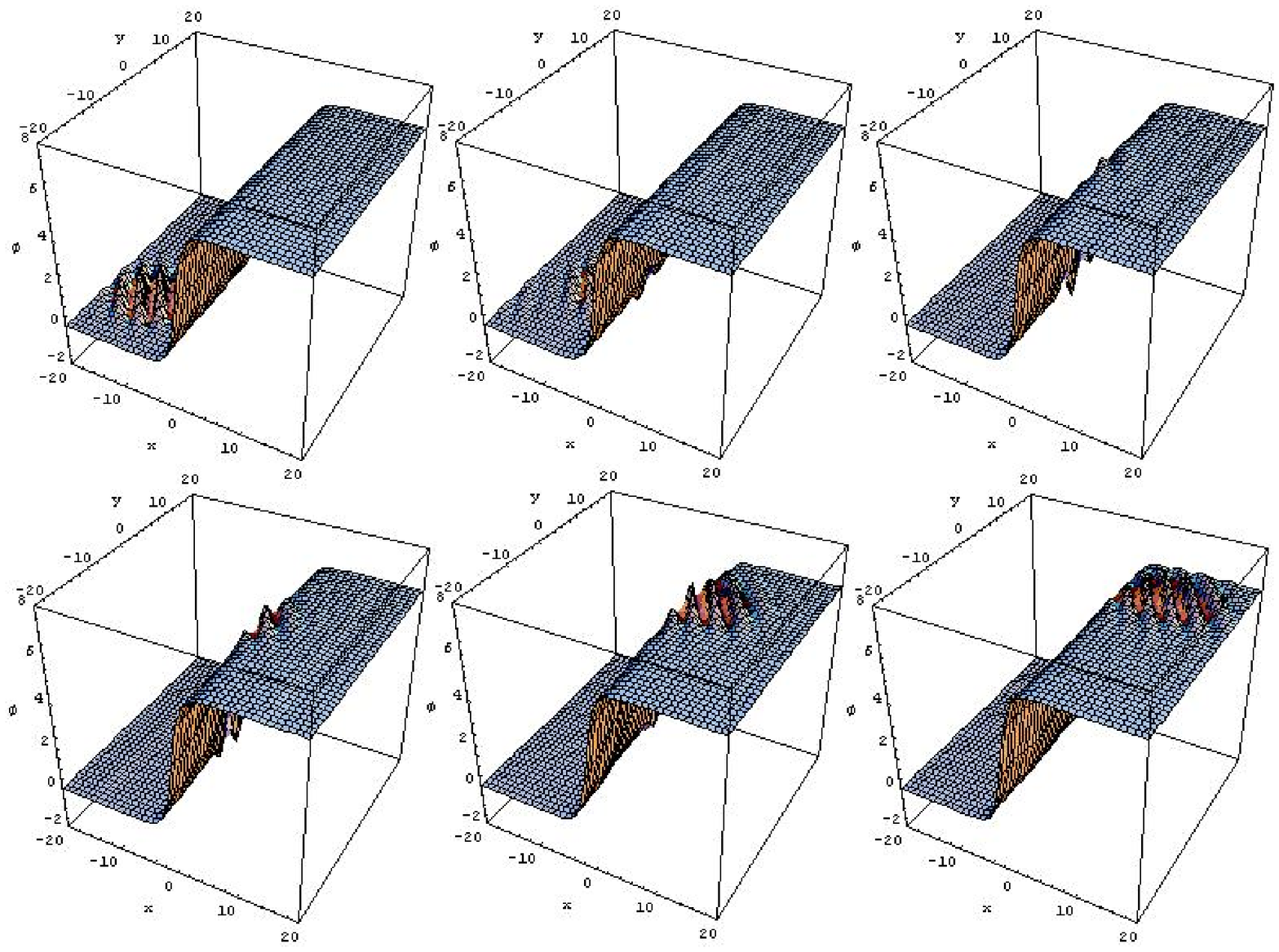}
\caption{Standing kink-light bullet 30$^{\circ}$-collision. The light bullet has $\left\{\gamma,\sigma\right\}=\left\{2.0,2.5\right\}$, $\stackrel{\rightarrow}{r_{01}}=(-15cos30^{\circ},-15sin30^{\circ})$ and $\stackrel{\rightarrow}{d_1}=cos30^{\circ}\stackrel{\rightarrow}{e_x}+sin30^{\circ}\stackrel{\rightarrow}{e_y}$, while the standing kink ($m=1$) presents $\stackrel{\rightarrow}{r_{02}}=\stackrel{\rightarrow}{0}$ and $\stackrel{\rightarrow}{d_2}=\stackrel{\rightarrow}{e_x}$. An $800\cdot61\cdot61$-point grid is defined in $\left\{\left(t,x,y\right): t\epsilon\left[0,100\right] \wedge x\epsilon\left[-30,30\right] \wedge y\epsilon\left[-30,30\right] \right\}$ and this domain is zoomed into $\left\{\left(t,x,y\right): t\epsilon\left[0,100\right] \wedge x\epsilon\left[-20,20\right] \wedge y\epsilon\left[-20,20\right] \right\}$. In this case, $\Gamma=3.14$.}
\label{figLBSK}
\end{figure}

 Moreover, a 30$^{\circ}$-collision between a light bullet and a standing kink is simulated - see figure \ref{figLBSK}. As in the previous scenarios, no modification in the light bullet propagation direction is detected, which evidences the robustness of these objects.

\par
\begin{figure}[h]
\centering 
\includegraphics[scale=0.43]{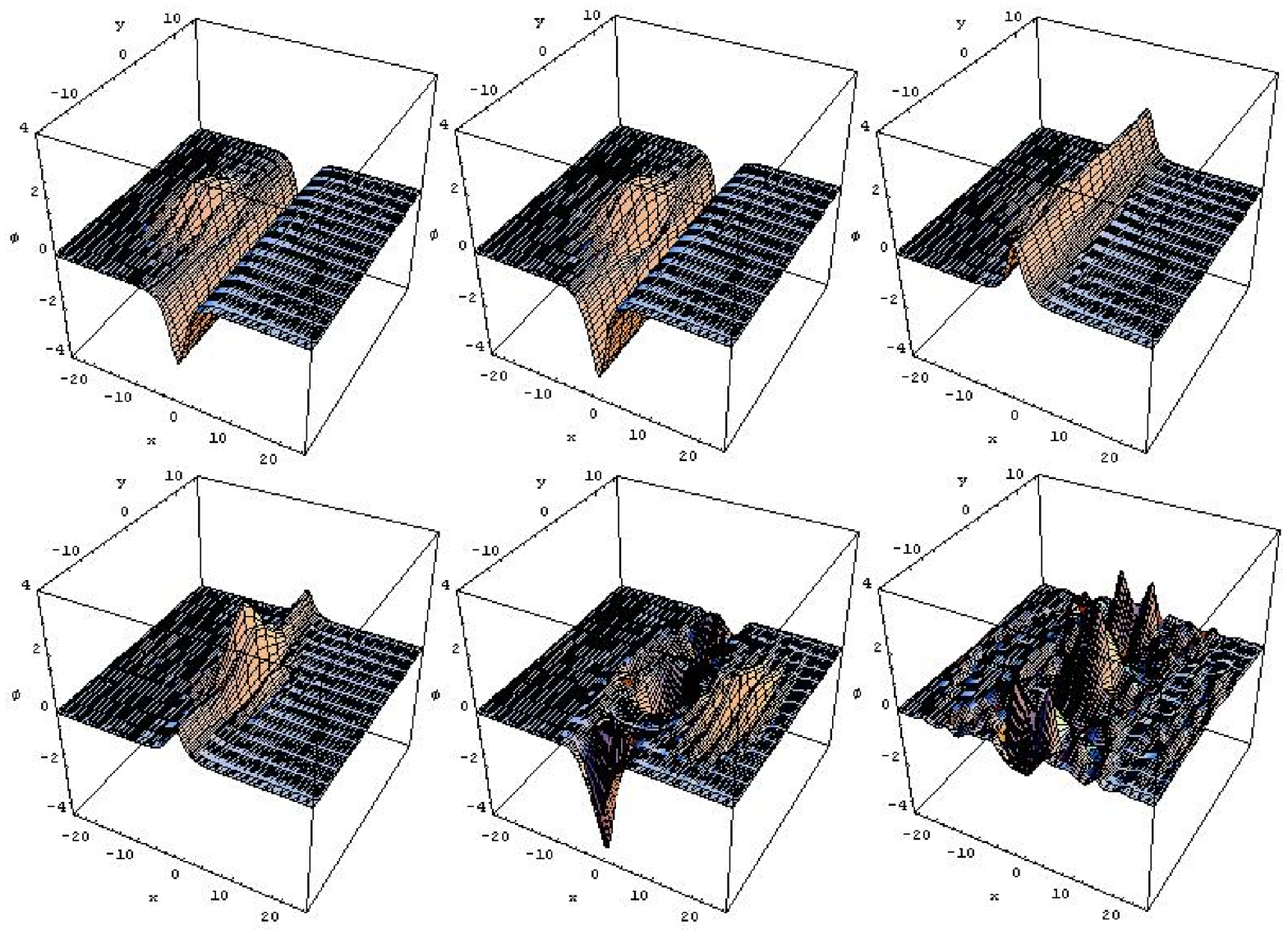}
\caption{Standing breather-light bullet head-on collision. The light bullet has $\left\{\gamma,\sigma\right\}=\left\{2.0,2.5\right\}$, $\stackrel{\rightarrow}{r_{01}}=(-10,0)$ and $\stackrel{\rightarrow}{d_1}=\stackrel{\rightarrow}{e_x}$, while the standing breather presents $m=0.8$, $\stackrel{\rightarrow}{r_{02}}=\stackrel{\rightarrow}{0}$ and $\stackrel{\rightarrow}{d_2}=\stackrel{\rightarrow}{e_x}$. An $800\cdot251\cdot31$-point grid is defined in $\left\{\left(t,x,y\right): t\epsilon\left[0,100\right] \wedge x\epsilon\left[-30,30\right] \wedge y\epsilon\left[-35,35\right] \right\}$ and this domain is zoomed into $\left\{\left(t,x,y\right): t\epsilon\left[0,100\right] \wedge x\epsilon\left[-25,25\right] \wedge y\epsilon\left[-15,15\right] \right\}$. In this case, $\Gamma=13.09$.}
\label{fig16}
\end{figure}

 Another collision taken into account is the one between a light bullet and a \textit{standing breather}, shown in figure \ref{fig16}. The latter object is an analytical solution of \eqref{eqSG} which is oscillatory and may be interpreted as a bound state between a kink and an antikink. Its behaviour is previewed analytically in the (1+1)-dimension case \cite{Infeld} and can be extended to
\begin{equation}\label{stdbreather}
\phi\left(t,\stackrel{\rightarrow}{r}\right)=4 arctg\left(\frac{m}{\sqrt{1-m^2}}\frac{sin\omega t}{cosh\left(m\stackrel{\rightarrow}{d} \cdot \left(\stackrel{\rightarrow}{r}-\stackrel{\rightarrow}{r_0}\right)\right)}\right)
\end{equation}
where $\omega=\frac{2\pi}{T}=\sqrt{1-m^2}$ and $|m|<1$. In this collision, the light bullet still emerges intact, although the standing breather is completely ruined after the interaction. In other words, the bound state kink-antikink is destroyed by the light bullet, even though this does not destroy isolated kinks nor antikinks.

\par Figures \ref{fig13}, \ref{fig14}, \ref{fig15}, \ref{figLBSK} and \ref{fig16} all together allow one to confirm the robustness of light bullets observed in \ref{32}. Indeed, these solutions pass through analytical solutions \eqref{1sol} and \eqref{stdbreather} and remain unchanged. This property may have interesting consequences, specially if one identifies light bullets with optical pulses propagating in different media \cite{Xin}.

\section{Final remarks}\label{4}
\par The analysis of kink and antikink solitons revealed an obvious shape consistency during propagation, which is typical of solitons behaviour. Moreover, an $m$-dependent study of the velocity in the kink case was carried out and it followed the theoretical solution \eqref{1sol}, as expected.

\par As for kink-kink collision, one verified that kinks do not alter their shape, propagation direction nor velocity after head-on collisions with each other. Nevertheless, the numerical value of $x_0$ was found to be significantly above 0 for both kinks, which proves that nonlinear interaction has taken place and that it momentarily speeded up both solitons.

\par In the light bullet case, the propagation was characterised as a function of the parameters $\gamma$ and $\sigma$ and the most favourable case was identified: $\left\{\gamma,\sigma\right\}=\left\{2.0,2.5\right\}$. With this configuration, a head-on collision was set up and nonlinear effects in the shape of the light bullets were measured (through $a_2$). The shape consistency of the light bullets noticed in the latter collision was also seen in head-on collisions with kink, antikink, standing kink and standing breather. 

\par Moreover, a 30$^{\circ}$-collision between a light bullet and a standing kink was set up as well and it reinforced the idea that light bullets and kinks are indeed robust objects. We also found that the direction of the propagation of the light bullet is maintained after it collides with the kink.

\par Future developments may study the dependence on $\gamma$ and $\sigma$ of light bullet collisions and verify if these lead to a shrink of the light bullets or not. Another interesting point is to understand why kink, antikink and standing kink survive to the collision with a light bullet while the standing breather does not.









\end{document}